\documentclass[aps,float,twocolumn,epsfig]{revtex4}
\usepackage{amsmath}
\usepackage{epsfig}

\begin{document}

\title{Resonant nonlinear optics in coherently prepared media: full analytic
solutions}
\author{E.A. Korsunsky and M. Fleischhauer}

\affiliation{Fachbereich Physik, Universit\"{a}t Kaiserslautern,
D-67663 Kaiserslautern, Germany}

\date{\today{}}

\begin{abstract}
We derive analytic solution for pulsed frequency conversion based on
electromagnetically induced transparency (EIT) or maximum coherence in
resonant atomic vapors. In particular drive-field and coherence depletion
are taken into account. The solutions are obtained with the help of an
Hamiltonian approach which in the adiabatic limit allows to reduce the full
set of Maxwell-Bloch equations to simple canonical equations of Hamiltonian
mechanics for the field variables. Adiabatic integrals of motion can be
obtained and general expressions for the spatio-temporal evolution of field
intensities derived. Optimum conditions for maximum conversion efficiency
are identified and the physical mechanism of nonlinear conversion in the
limit of drive-field and coherence depletion discussed.
\end{abstract}

\maketitle

\section{Introduction}

Resonant nonlinear optics in atomic gases has received a great impetus in
recent years due to new concepts based on the application of specific
coherence and interference effects. Two of these concepts attracted
particular attention. The first mechanism uses the effect of
electromagnetically induced transparency (EIT) \cite{har97,harr90}, the
second maximum coherence between two metastable atomic levels \cite{jain96}.
Both schemes allow for high-efficiency nonlinear conversion with
substantially alleviated phase-matching problems in a rather dilute ensemble
of atoms if the medium is driven by an undepleted coherent coupling field or
if an undepleted atomic coherence is assumed. These assumptions do not take
into account the resources needed to maintain the drive field or atomic
coherence. Thus in the considered limit the \textit{overall} efficiency of
the resonant nonlinear processes is small. In the present paper we analyze
and compare EIT- and maximum-coherence based systems with each other and
with the conventional non-resonant schemes of nonlinear optics taking into
account the transparency- or coherence-maintaining fields and their
depletion. Effects of inhomogeneous broadening are however disregarded.
Deriving analytic solutions for the nonlinear pulse propagation problem, we
show that in both systems also a large \textit{overall} conversion
efficiency can be achieved. The physical nature of the conversion can
however no longer be associated with EIT or maximum coherence alone.

In EIT a strong resonant electromagnetic (\textit{em}) field of frequency $%
\omega _{2}$ is applied to the transition between two excited states $\left|
2\right\rangle $ and $\left| 3\right\rangle $, (Fig.1a), causing a splitting
of both (Autler-Townes effect). When a weak probe field of frequency $\omega
_{3}$ resonant with the transition from the ground-state $|1\rangle $ to the
bare state $|3\rangle $ is applied, its linear interaction with the medium
is almost perfectly cancelled. The vanishing of the linear susceptibility,
i.e. absorption and refraction, for the probe field is due to a destructive
interference between the two excitation paths through the Autler-Townes
doublet and persists even for small splittings. If the ground state $%
|1\rangle $ is coupled to the intermediate state $|2\rangle $, on the other
hand (e.g. by a two-photon transition), photons of frequency $\omega _{3}$
are generated. While the resonant linear absorption of these photons is
suppressed by destructive quantum interference (EIT), the nonlinear
susceptibility responsible for the generation of them is only affected by
the splitting, which can be rather small. Since there are no resonant
contributions to the refractive index, there is furthermore perfect phase
matching except for contributions from other off-resonant transitions. Owing
to the large dispersion near the transparency frequency of EIT, a small
detuning $\delta _{2}$ can be applied to compensate for these contributions
and phase matching can easily be obtained without sacrificing the
cancellation of absorption. Thus EIT provides a perfect system for large
nonlinear conversion with a minimum of atoms. Since the first demonstration
of phase matching in EIT-assisted four-wave frequency mixing \cite{jain93},
many other experiments confirmed considerable improvement in the conversion
efficiency when EIT has been used (see, e.g., Refs. \cite{stoi,maran}).

Another mechanism proposed recently is referred to as nonlinear optics with
maximum coherence \cite{jain96,mer99}. The idea is here to prepare and
maintain the atoms in a coherent superposition of atomic states $\left|
1\right\rangle $ and $\left| 2\right\rangle $ with equal amplitudes. This
can be done either by two strong fields exciting the $\left| 1\right\rangle
-\left| 2\right\rangle $ coherence via a Raman transition \cite{jain96} or
by other means such as rapid adiabatic passage \cite{scrap}. The coherence,
established in such a way, plays the role of a strong local oscillator. If
then a relatively weak field $\omega _{2}$ detuned by a sufficiently large
amount $\delta _{3}$ from the $\left| 2\right\rangle -\left| 3\right\rangle $
transition is applied, Fig. 1b, it will beat against the local oscillator to
generate the sum (or difference) frequency. Here, the refraction due to
resonant transitions does not vanish. But since the nonlinear coupling
strength is of the same order as the linear susceptibility, complete
conversion occurs over a distance smaller than the coherence length \cite
{boyd}. Consequently there is no need to phasematch the propagating beams.
We note that the relation between the $\left| 1\right\rangle -\left|
2\right\rangle $ coherence and the wave mixing processes has been pointed
out in earlier work on nonlinear optics (see, e.g., Refs. \cite{wil82} and
references therein).

In the limit of undepleted drive or constant coherence, the nonlinear
conversion process in both the EIT and maximum coherence scheme affects the
quantum state of the atoms only in the perturbative sense. Consequently the
probability of photon conversion per atom is small. For the same reason the
overall conversion efficiency including the transparency and coherence
generating fields is small. We here analyze the properties of the two
mechanisms for arbitrary strength of the drive fields taking into account
their depletion and compare them to conventional off-resonant nonlinear
optics.

The drive depletion turns the propagation problem into a truly nonlinear
one. Its solution is particularly challenging for pulses and is in general
possible only numerically. In order to obtain transparent analytical
solutions we here apply the so-called Hamiltonian approach \cite
{mel79,kryz,McK90,cap91} which allows for a solution in a wide range of
physically relevant situations. This approach is especially useful under
adiabatic conditions, i.e. when the atoms are excited by the laser pulses in
such a way that they remain in the same instantaneous eigenstate of the
interaction Hamiltonian during the entire process. In the present work, we
will assume that the interaction is adiabatic. The adiabatic approximation
requires a slow rate of evolution as compared to the frequency separation of
the adiabatic eigenstates. This results usually in a requirement for the
product of the pulse duration and the Rabi frequency of the radiation field
to be much larger than unity. Thus, for sufficiently intense fields, the
process can be adiabatic even for short pulses. It should be noted however
that this assumption rules out such important effects as group velocity
reduction which results from lowest-order non-adiabatic corrections. For
sufficiently large intensities or cw fields these effects do not influence
the process. An analysis of non-adiabatic corrections to the Hamiltonian
approach taking into account group delays will be given in a future
publication.

To simplify the calculations and the interpretation of the analytic results,
we restrict ourselves to a somewhat idealized three-level atomic excitation
scheme (Fig. 1). The nonlinear three-wave mixing in a three-level system is
normally forbidden due to symmetry \cite{boyd}. However, this scheme
provides the simplest example where all above mentioned mechanisms of
nonlinear-optical wave mixing may take place. It should be noted that
similar results, however with much lengthier expressions and larger number
of atomic parameters, can be found for a more realistic three-level scheme
where $\left| 1\right\rangle -\left| 2\right\rangle $ transition is a
two-photon one \cite{to-be-published}. Moreover, the 3-wave-mixing processes
is possible when a dc electric field is applied to the atomic sample \cite
{hak91}.

The paper is organized as follows. In Sec.II we discuss resonant nonlinear
optical processes based on EIT or maximum coherence for undepleted drive
fields or undepleted coherence, respectively, and compare them to
conventional off-resonance nonlinear optics. In Sec.III we outline the
Hamiltonian approach, which allows to eliminate the atomic degrees of
freedom assuming adiabatic following and to map the pulse propagation
problem to the dynamics of a one-dimensional nonlinear pendulum. Making use
of this formalism we derive full analytic solution for EIT and
maximum-coherence based nonlinear optics in Sec.IV, taking into account
drive-field and coherence depletion.


\section{Resonant nonlinear optics in a coherently prepared 3-level system}


We first consider the propagation of pulsed \textit{em} fields in a medium
of three-level atoms (Fig. 1) in which either a constant drive field mixes
the two excited states $|2\rangle $ and $|3\rangle $ or in which a constant
coherence between the lower two states $|1\rangle $ and $|2\rangle $ is
maintained. The first case corresponds to resonant nonlinear frequency
conversion based on EIT \cite{har97,harr90}, the second one to nonlinear
optics with maximum coherence \cite{jain96}. The electric field propagating
in the $z$-direction is assumed to consist of three components with carrier
frequencies $\omega _{1},\omega _{2}$ and $\omega _{3}=\omega _{1}+\omega
_{2}$:
\begin{equation}
E(z,t)=\sum_{j}\Bigl(\mathcal{E}_{j}(z,t)\exp (-i(\omega _{j}t-{k}%
_{j}z))+c.c.\Bigr).  \label{E}
\end{equation}
Here ${k}_{j}=n_{j}\omega _{j}/c$\ with $n_{j}$ being the refractive index
at frequency $\omega _{j}$ due to levels outside the three-level system of
Fig. 1. This background refraction gives rise to the ''residual'' phase
mismatch determined as:
\begin{equation}
\Delta k=k_{1}+k_{2}-k_{3}.  \label{dk}
\end{equation}


\begin{figure}[th]
\centerline{\epsfig{file=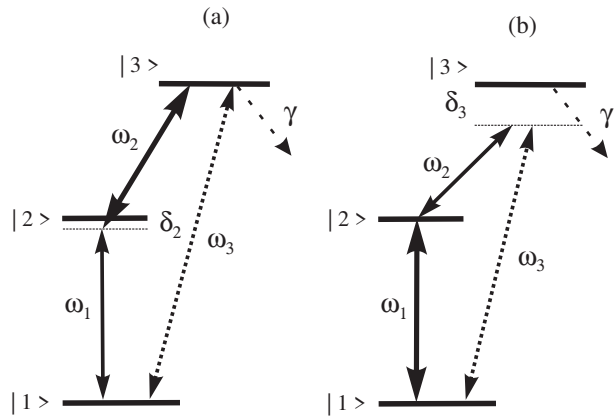,width=8.0 cm}} \vspace*{2ex}
\caption{Resonant sum-frequency generation. (a) strong drive field ($\protect%
\omega _{2}$) between metastable state $|2\rangle $ and excited state $%
|3\rangle $ creates EIT; (b) strong drive $\protect\omega _{1}$ (two-photon,
Raman or magnetic coupling) generates maximum coherence between $|1\rangle $
and $|2\rangle $.}
\label{system}
\end{figure}


In the approximation of slowly varying amplitudes and phases Maxwell's
propagation equations read in a moving frame
\begin{equation}
\frac{\partial \mathcal{E}_{j}}{\partial z}=i2\pi \frac{\omega _{j}}{c}%
\mathcal{P}_{j},  \label{Max1}
\end{equation}
where $\mathcal{E}_{j}$ and $\mathcal{P}_{j}$ are functions of the
coordinate $z$ and the retarded time $\tau =t-z/c$. $\mathcal{P}_{j}$ are
the components of the medium polarization:
\begin{equation*}
P=\sum_{j}\Bigl(\mathcal{P}_{j}\exp (-i(\omega _{j}t-k_{j}z))+c.c.\Bigr),
\end{equation*}
which can be expressed in terms of the atomic probability amplitudes $c_{n}$
in levels $|1\rangle $, $|2\rangle $ and $|3\rangle $:
\begin{eqnarray}
\mathcal{P}_{1} &=&N\,d_{1}c_{1}^{\ast }c_{2},  \label{P1} \\
\mathcal{P}_{2} &=&N\,d_{2}\,c_{2}^{\ast }c_{3}\mathrm{e}^{i\theta -i\Delta
kz},  \label{P2} \\
\mathcal{P}_{3} &=&N\,d_{3}c_{1}^{\ast }c_{3},  \label{P3}
\end{eqnarray}
$N$ being the density of active atoms. $d_{1}$, $d_{2}$ and $d_{3}$ are the
real dipole moments of the transitions $|1\rangle \rightarrow |2\rangle $, $%
|2\rangle \rightarrow |3\rangle $ and $|1\rangle \rightarrow |3\rangle $
respectively. $\theta $ is phase of the $|2\rangle \rightarrow |3\rangle $
dipole moment, which in general cannot be chosen freely.

Assuming decay only from the top most state $|3\rangle $ with rate $\gamma $
the state amplitudes obey in rotating wave approximation the equations
\begin{eqnarray}
\dot{c}_{1} &=&i\Omega _{1}^{\ast }\,c_{2}+i\Omega _{3}^{\ast }\,c_{3},
\notag \\
\dot{c}_{2} &=&i\Omega _{1}\,c_{1}+i\Omega _{2}^{\ast }\,\mathrm{e}^{i\theta
-i\Delta kz}\,c_{3}+i\delta _{2}c_{2},  \label{ampl} \\
\dot{c}_{3} &=&i\Omega _{2}\,\mathrm{e}^{-i\theta +i\Delta kz}c_{2}+i\Omega
_{3}c_{1}+i\left( \delta _{3}+i\gamma \right) c_{3},  \notag
\end{eqnarray}
where $\delta _{2}$ and $\delta _{3}$ are the frequency detunings indicated
in Fig.~1.
\begin{equation}
\delta _{2}=\omega _{1}-\omega _{21},\qquad \delta _{3}=\omega _{3}-\omega
_{31},  \label{d3}
\end{equation}
with $\omega _{ij}$ denoting the transition frequencies between the
corresponding levels. $\Omega _{1}$, $\Omega _{2}$ and $\Omega _{3}$ are the
Rabi frequencies for transitions $|1\rangle -|2\rangle $, $\left|
2\right\rangle -\left| 3\right\rangle $ and $\left| 1\right\rangle -\left|
3\right\rangle $, respectively:
\begin{equation}
\Omega _{j}=\frac{d_{j}\mathcal{E}_{j}}{2\hbar }.
\end{equation}


\subsection{EIT with undepleted coupling field}


Here, we consider three fields interacting with the 3-level system. For
simplicity we assume $\delta _{3}=0$. Furthermore we consider the case of a
strong, undepleted drive field with frequency $\omega _{2}$
\begin{equation*}
\left| \Omega _{2}\right| \gg \left| \Omega _{1}\right| ,\left| \Omega
_{3}\right| ,\gamma ,\left| \delta _{2}\right| .
\end{equation*}
Then, the solution of the atomic equations of motion (\ref{ampl}) with the
initial condition $c_{1}\left( t\rightarrow -\infty \right) =1$ yields
\begin{equation*}
\left| c_{1}\right| \approx 1.
\end{equation*}
Assuming quasi-adiabatic evolution, i.e. not too fast changing fields, we
find
\begin{eqnarray*}
c_{2} &=&-\frac{\Omega _{3}\Omega _{2}^{\ast }e^{i\theta -\Delta kz}-i\Omega
_{1}\gamma }{\left| \Omega _{2}\right| ^{2}-i\delta _{2}\gamma }, \\
c_{3} &=&-\frac{\Omega _{1}\Omega _{2}e^{-i\theta +i\Delta kz}-\Omega
_{3}\delta _{2}}{\left| \Omega _{2}\right| ^{2}-i\delta _{2}\gamma }.
\end{eqnarray*}
Substitution into Maxwell's propagation equations gives:
\begin{eqnarray}
\frac{\partial \mathcal{E}_{1}}{\partial z} &=&-\frac{\pi Nd_{1}^{2}}{\hbar }%
\frac{\omega _{1}}{c}\frac{\gamma }{\left| \Omega _{2}\right| ^{2}}\mathcal{E%
}_{1}  \label{e1e} \\
&&-i\frac{\pi Nd_{1}d_{3}}{\hbar \left| \Omega _{2}\right| }\frac{\omega _{1}%
}{c}e^{i\theta -i\Delta kz}\mathcal{E}_{3},  \notag \\
\frac{\partial \mathcal{E}_{3}}{\partial z} &=&i\frac{\pi Nd_{3}^{2}}{\hbar }%
\frac{\omega _{3}}{c}\frac{\delta _{2}}{\left| \Omega _{2}\right| ^{2}}%
\mathcal{E}_{3}  \label{e3e} \\
&&-i\frac{\pi Nd_{1}d_{3}}{\hbar \left| \Omega _{2}\right| }\frac{\omega _{3}%
}{c}e^{-i\theta +i\Delta kz}\mathcal{E}_{1}.  \notag
\end{eqnarray}
Equations (\ref{e1e}) and (\ref{e3e}) are linear differential equations,
which can easily be solved. We now consider the case in which no $\mathcal{E}%
_{3}$ field is incident on the medium, $\mathcal{E}_{3}\left( z=0\right) =0$%
. Introducing the normalized intensity (photon flux)
\begin{equation}
\eta _{j}=\frac{{I_{j}}}{{\hbar \omega _{j}}}\equiv \frac{c\left| \mathcal{E}%
_{j}\right| ^{2}}{8\pi \hbar \omega _{j}}  \label{eta}
\end{equation}
and the coupling strength
\begin{equation}
\mu _{j}=\frac{2\pi \omega _{j}d_{j}^{2}}{\hbar c},  \label{mus}
\end{equation}
the solution of equations (\ref{e1e}) and (\ref{e3e}) reads:
\begin{equation}
\eta _{3}(z)=\frac{\eta _{10}}{1+\left( \Delta k^{\prime }/2\kappa \right)
^{2}}\,\mathrm{e}^{-\Gamma \kappa z}\,\sin ^{2}\left( \kappa z\sqrt{1+\left(
\Delta k^{\prime }/2\kappa \right) ^{2}}\right) ,  \label{sol1}
\end{equation}
where $\eta _{10}=\eta _{1}(z=0)$ is the photon flux at the entrance to the
medium and we have introduced the conversion coefficient $\kappa $:
\begin{equation}
\kappa =\frac{N}{2}\frac{\sqrt{\mu _{1}\mu _{3}}}{|\Omega _{2}|},
\label{kap-e}
\end{equation}
the loss coefficient $\Gamma $:
\begin{equation}
\Gamma =\frac{\gamma }{|\Omega _{2}|}\sqrt{\frac{\mu _{1}}{\mu _{3}}},
\label{G-e}
\end{equation}
and
\begin{equation}
\Delta k^{\prime }=\Delta k-\frac{N}{2}\frac{\mu _{3}\delta _{2}}{\left|
\Omega _{2}\right| ^{2}}  \label{dk-e}
\end{equation}
is the total phase mismatch, including the background value $\Delta k$ and
the contribution from resonant transition $|1\rangle \rightarrow |3\rangle $%
. In the EIT case, the resonant contribution to the phase mismatch is always
smaller than the conversion coefficient $\kappa $ by a factor $\sim \delta
_{2}/|\Omega _{2}|$. Moreover, a small detuning $\delta _{2}$ can be
introduced to compensate the residual phase mismatch. Thus, the EIT scheme
represents an ideal situation with complete phase matching, where the
optimum conversion occurs for a density-length product
\begin{equation}
N\,z\bigr\vert_{\mathrm{opt}}=\pi \frac{|\Omega _{2}|}{\sqrt{\mu _{1}\mu _{3}%
}}.
\end{equation}
In order to minimize the absorption losses, Eq. (\ref{G-e}), the Rabi
frequency of the coupling field $\left| \Omega _{2}\right| $ has to be
sufficiently large:
\begin{equation*}
|\Omega _{2}|\gg \sqrt{\frac{\mu _{1}}{\mu _{3}}}\gamma .
\end{equation*}


\subsection{Undepleted coherence $\protect\rho _{12}$}


In the following we discuss the scheme of resonant nonlinear optics with
maximum coherence. For this situation, constant state amplitudes $c_{1}$ and
$c_{2}$ are assumed, maintained e.g. by a constant strong drive field $%
\Omega _{1}$. In this case the wave vector of the atomic coherence $%
c_{1}^{\ast }c_{2}$ is equal to $k_{1}$ which can be changed by, e.g., the
small detuning $\delta _{2}$ \cite{jain96}. The backaction of the atoms to
this field is disregarded and thus the corresponding coupling does not need
to be taken into account. Under the condition
\begin{equation*}
\left| \delta _{3}+i\gamma \right| \gg \left| \Omega _{j}\right|
\end{equation*}
the amplitude of the excited state can be adiabatically eliminated, which
yields
\begin{equation}
c_{3}=-\frac{\Omega _{2}e^{-i\theta +i\Delta kz}c_{2}+\Omega _{3}c_{1}}{%
\delta _{3}+i\gamma }.  \label{c3}
\end{equation}
Substitution of Eqs. (\ref{P2}), (\ref{P3}), (\ref{c3}) into the Maxwell
propagation equations (\ref{Max1}) gives:
\begin{eqnarray}
\frac{\partial \mathcal{E}_{2}}{\partial z} &=&-i\frac{\pi Nd_{2}^{2}}{\hbar
}\frac{\omega _{2}}{c}\frac{\delta _{3}-i\gamma }{\delta _{3}^{2}+\gamma ^{2}%
}\left| c_{2}\right| ^{2}\mathcal{E}_{2}  \label{e21} \\
&&-i\frac{\pi Nd_{2}d_{3}}{\hbar }\frac{\omega _{2}}{c}\frac{\delta
_{3}-i\gamma }{\delta _{3}^{2}+\gamma ^{2}}\rho _{12}e^{i\theta -i\Delta kz}%
\mathcal{E}_{3},  \notag \\
\frac{\partial \mathcal{E}_{3}}{\partial z} &=&-i\frac{\pi Nd_{3}^{2}}{\hbar
}\frac{\omega _{3}}{c}\frac{\delta _{3}-i\gamma }{\delta _{3}^{2}+\gamma ^{2}%
}\left| c_{1}\right| ^{2}\mathcal{E}_{3}  \label{e31} \\
&&-i\frac{\pi Nd_{2}d_{3}}{\hbar }\frac{\omega _{3}}{c}\frac{\delta
_{3}-i\gamma }{\delta _{3}^{2}+\gamma ^{2}}\rho _{12}e^{-i\theta +i\Delta kz}%
\mathcal{E}_{2},  \notag
\end{eqnarray}
where $\rho _{12}=\left| c_{1}^{\ast }c_{2}\right| $.

This is again a set of linear differential equations whose solution has the
same form as that for the EIT case, Eq. (\ref{sol1}), with the substitution $%
\eta _{10}\rightarrow \eta _{20}$ and corresponding parameters (assuming $%
\delta _{3}\gg \gamma $):
\begin{eqnarray}
\kappa &=&\frac{N}{2}\frac{\sqrt{\mu _{2}\mu _{3}}}{\delta _{3}}\rho _{12},
\label{kap-m} \\
\Gamma &=&\frac{\gamma }{\delta _{3}}\frac{\mu _{2}\left| c_{2}\right|
^{2}+\mu _{3}\left| c_{1}\right| ^{2}}{\rho _{12}\sqrt{\mu _{2}\mu _{3}}},
\label{G-m} \\
\Delta k^{\prime } &=&\Delta k+\frac{N}{2}\frac{\mu _{3}\left| c_{1}\right|
^{2}-\mu _{2}\left| c_{2}\right| ^{2}}{\delta _{3}}.  \label{dk-m}
\end{eqnarray}
The total phase mismatch $\Delta k^{\prime }$ includes the background value $%
\Delta k$ and the contributions from resonant transitions $|1\rangle
\rightarrow |3\rangle $ and $|2\rangle \rightarrow |3\rangle $.

For atomic media which we consider here, the off-resonant (background)
contributions to the refractive index $n_{j}$ are of the order of \cite{boyd}%
:
\begin{equation}
n_{j}\approx 1+N\frac{c}{\omega _{j}}\sum_{m}\frac{\mu _{jm}}{\delta _{jm}},
\label{n}
\end{equation}
where $\mu _{jm}$ and $\delta _{jm}$ are the coupling constants and
detunings, respectively, for the wave with frequency $\omega _{j}$ and
transition to the far-detuned state $\left| m\right\rangle $ not belonging
to the three-level system of Fig. 1. Since $\delta _{jm}\gg \delta _{3}$,
the resonant contributions to the phase mismatch are the dominant ones. As
can be seen from Eqs.(\ref{kap-m}) and (\ref{dk-m}), they are in general of
the same order as the conversion coefficient $\kappa $ if the atomic
coherence $\rho _{12}$ is large (of the order of $1/2$). Therefore,
efficient energy transfer from the $\omega _{2}$ field into the $\omega _{3}$
field: $\eta _{3}(l)\sim \eta _{20}$, occurs already within a length $%
l=\kappa ^{-1}\sim L_{c}=2/\Delta k^{\prime },$ the coherence length. This
feature constitutes the main advantage of the maximum coherence scheme over
conventional nonlinear optics, because the phase matching is no longer
important.

If the conversion length $l$ is much smaller than the coherence length $L_{c}
$, $l/L_{c}\ll 1$, and the losses are small, $\left( \Gamma /l\right) z|_{%
\mathrm{opt}}\ll 1$, maximum conversion occurs for a density-length product
\begin{equation}
\,N\,z\bigr\vert_{\mathrm{opt}}=\frac{\pi }{2}Nl=\pi \frac{\delta _{3}}{\rho
_{12}\sqrt{\mu _{2}\mu _{3}}}.
\end{equation}
$l/L_{c}\ll 1$ is realized when the parameters are chosen such that $\Delta
k^{\prime }$ is small, or even better, vanishes, i.e. if
\begin{equation*}
\frac{2\Delta k}{N}\approx \frac{\mu _{2}\left| c_{2}\right| ^{2}-\mu
_{3}\left| c_{1}\right| ^{2}}{\delta _{3}}.
\end{equation*}
This can be achieved in different ways: (i) by tuning the wave vector $k_{1}$
of the atomic coherence (e.g., by introducing a detuning $\delta _{2}$, as
in Ref. \cite{jain96}), (ii) by selecting the appropriate detuning $\delta
_{3}$, as in Ref. \cite{mer99}, and/or, by preparation of atoms in a
superposition with suitable amplitudes $c_{1},c_{2}$.

In order for the absorption losses to be negligible within the optimum
propagation distance, the following condition has to be satisfied:
\begin{equation*}
\frac{\gamma }{\delta _{3}}\frac{1}{\rho _{12}}\ll 1,
\end{equation*}
which indicates once again that it is indeed advantageous to prepare a large
atomic coherence $\rho _{12}$.


\subsection{Conventional nonlinear optics: weak excitation}


Now, we discuss the case of a weak excitation of atoms which corresponds to
the regime of conventional nonlinear optics. The weak excitation takes
place, for example, when both detunings are very large:

\begin{equation*}
\left| \delta _{3}\right| ,\left| \delta _{2}\right| \gg \left| \Omega
_{j}\right| ,\gamma .
\end{equation*}
Under this condition, the atomic probability amplitudes are:
\begin{eqnarray*}
\left| c_{1}\right| &\approx &1, \\
c_{2} &=&-\frac{\Omega _{1}\Omega _{2}e^{-i\theta +i\Delta kz}-\Omega
_{3}\delta _{2}}{\delta _{2}\left( \delta _{3}+i\gamma \right) }, \\
c_{3} &=&-\frac{\Omega _{3}\Omega _{2}^{\ast }e^{i\theta -i\Delta kz}-\Omega
_{1}\left( \delta _{3}+i\gamma \right) }{\delta _{2}\left( \delta
_{3}+i\gamma \right) }.
\end{eqnarray*}
Since $\left| c_{2}\right| ,\left| c_{3}\right| \ll \left| c_{1}\right| $,
the medium polarization at frequency $\omega _{2}$ is much smaller than for
the $\omega _{1}$ and $\omega _{3}$ components: $\left| \mathcal{P}%
_{2}\right| \ll \left| \mathcal{P}_{1}\right| ,\left| \mathcal{P}_{3}\right|
$. Therefore, we can assume that the $\omega _{2}$ field is almost
undepleted, $\left| \Omega _{2}\right| \approx const(z)$.

The propagation equations read for this case:
\begin{eqnarray}
\frac{\partial \mathcal{E}_{1}}{\partial z} &=&-i\frac{\pi Nd_{1}^{2}}{\hbar
\delta _{2}}\frac{\omega _{1}}{c}\mathcal{E}_{1}  \label{e1n} \\
&&+i\frac{\pi Nd_{1}d_{3}\omega _{1}}{\hbar c}\frac{\left| \Omega
_{2}\right| }{\delta _{2}\delta _{3}}\left( 1-i\frac{\gamma }{\delta _{3}}%
\right) e^{i\theta -i\Delta kz}\mathcal{E}_{3},  \notag \\
\frac{\partial \mathcal{E}_{3}}{\partial z} &=&-i\frac{\pi Nd_{3}^{2}}{\hbar
}\frac{\omega _{3}}{c}\frac{\delta _{3}-i\gamma }{\delta _{3}^{2}+\gamma ^{2}%
}\mathcal{E}_{3}  \label{e3n} \\
&&-i\frac{\pi Nd_{1}d_{3}\omega _{3}}{\hbar c}\frac{\left| \Omega
_{2}\right| }{\delta _{2}\delta _{3}}\left( 1-i\frac{\gamma }{\delta _{3}}%
\right) e^{-i\theta +i\Delta kz}\mathcal{E}_{1}.  \notag
\end{eqnarray}
The solution of this set of propagation equations has exactly the same form
as in the maximum-coherence and EIT cases, Eq. (\ref{sol1}), with
parameters:
\begin{eqnarray}
\kappa &=&\frac{N}{2}\frac{|\Omega _{2}|\sqrt{\mu _{1}\mu _{3}}}{\delta
_{2}\delta _{3}},  \label{kap-n} \\
\Gamma &=&\frac{\gamma }{\delta _{3}}\sqrt{\frac{\mu _{1}}{\mu _{3}}}\frac{%
\delta _{2}}{|\Omega _{2}|},  \label{G-n} \\
\Delta k^{\prime } &=&\Delta k+\frac{N}{2}\left( \frac{\mu _{3}}{\delta _{3}}%
-\frac{\mu _{1}}{\delta _{2}}\right) .  \label{Lc-n}
\end{eqnarray}
We see that the conversion length $l=\kappa ^{-1}$ is here much larger than
both in the maximum coherence, Eq. (\ref{kap-m}), and in the EIT, Eq. (\ref
{kap-e}), schemes. In fact, the conversion length is also much larger than
the (non-phase-matched) coherence length $L_{c}=2/\Delta k^{\prime }$ as $%
\delta _{2,3}/|\Omega _{2}|\gg 1$. Therefore, it is not possible in
conventional nonlinear optics to get the complete conversion, $\eta _{3}\sim
\eta _{10}$, without careful phase matching.

If such a compensation is performed, i.e. if $\left( \kappa /2\Delta
k^{\prime }\right) ^{2}\gg 1$ (by proper choice of detunings $\delta _{2}$
and $\delta _{3}$), optimum conversion occurs for a density-length product
\begin{equation}
N\,z\bigr\vert_{\mathrm{opt}}=\frac{\pi }{2}Nl=\pi \frac{\delta _{2}\delta
_{3}}{\sqrt{\mu _{1}\mu _{3}}\left| \Omega _{2}\right| }.
\end{equation}
At this optimum propagation distance, the relative absorption losses are
given by the parameter $\Gamma $, Eq. (\ref{G-n}).

Thus, we conclude that in the limit of undepleted drive field(s) both EIT
and maximum coherence schemes perform much better than conventional
nonlinear optics, regarding the number of atoms $N\,z\bigr\vert_{\mathrm{opt}%
}$ necessary for optimum conversion and robustness to phase mismatch. The
aim of the present paper is to investigate whether the \textit{em} energy
can be transferred to the generated wave from both pump fields, and whether
the attractive features of EIT- and maximum coherence-assisted conversion
survive when the drive field is depleted. This requires the solution of the
complete nonlinear propagation problem which we treat with the Hamiltonian
formalism outlined in the following section.


\section{Hamiltonian approach}


The resonant interaction of \textit{em} fields with atomic systems is
described by Maxwell's equations for the fields and a set of master
equations for the density matrix of the atoms. The solution of this coupled
set is rather difficult and except for some very special cases impossible
analytically. In order to derive analytical solutions for the field
propagation, approximations are needed that allow to eliminate the atomic
degrees of freedom and to express the polarization in terms of field
variables. A common approximation, which is not perturbative in the fields,
is the adiabatic solution of the density matrix equations. This
straightforward approach provides, when applicable, full information about
both atoms and fields in the adiabatic limit.


\subsection{General formalism}


Instead of the usual explicit scheme of adiabatic elimination which is
rather cumbersome, we use here a different, implicit approach \cite
{mel79,kryz} which yields directly effective field equations. This approach
is based on the representation of the medium polarization $P$ as a partial
derivative of the time-averaged free energy density of a dielectric with
respect to the electric field $E$ \cite{LL}:
\begin{equation}
P=-\left\langle N\frac{\partial \hat{H}}{\partial E}\right\rangle ,
\label{Pol2}
\end{equation}
where $\left\langle ...\right\rangle $ denotes quantum-mechanical averaging,
and $\hat{H}$ is the single-atom interaction Hamiltonian. For the field
given by Eq. (\ref{E}), we can write:
\begin{equation*}
P=-\left\langle N\sum_{j}\frac{\partial \hat{H}}{\partial \mathcal{E}%
_{j}^{\ast }}\exp (-i(\omega _{j}t-k_{j}z))+c.c.\right\rangle ,
\end{equation*}
so that the propagation equation becomes:
\begin{equation}
\frac{\partial \mathcal{E}_{j}}{\partial z}=-i2\pi \frac{\omega _{j}}{c}%
N\left\langle \frac{\partial \hat{H}}{\partial \mathcal{E}_{j}^{\ast }}%
\right\rangle .  \label{Max2}
\end{equation}
We here consider light-atom interaction processes that are adiabatic, that
is the atomic system can be assumed to follow the evolution of the
instantaneous eigenstates. For example, if the atomic system is at some
initial time $t_{0}$ in the nondegenerate eigenstate $\left| \psi
_{0}(t_{0})\right\rangle $ of the interaction Hamiltonian, i.e.
\begin{equation}
\hat{H}\left| \psi _{0}\right\rangle =\hbar \lambda _{0}\left| \psi
_{0}\right\rangle ,  \label{Eig1}
\end{equation}
which is at $t=t_{0}$ usually identical to the ground state of the
atoms, it will remain in this state at all times. Furthermore, we
disregard irreversible dissipation processes. As we show later,
for the processes we consider here, this is justified even for
rather long pulses, i.e., longer than the natural lifetime $\gamma
^{-1}$ of the excited state $\left| 3\right\rangle $. In this
limit one finds
\begin{equation*}
\left\langle \frac{\partial \hat{H}}{\partial \mathcal{E}_{j}^{\ast }}%
\right\rangle =\left\langle \psi _{0}\right| \frac{\partial \hat{H}}{%
\partial \mathcal{E}_{j}^{\ast }}\left| \psi _{0}\right\rangle =\hbar \frac{%
\partial \lambda _{0}}{\partial \mathcal{E}_{j}^{\ast }}.
\end{equation*}
Hence the propagation equation can be written as:
\begin{equation}
\frac{\partial \mathcal{E}_{j}}{\partial z}=-i2\pi \frac{\hbar \omega _{j}}{c%
}N\frac{\partial \lambda _{0}}{\partial \mathcal{E}_{j}^{\ast }}.
\label{Max3}
\end{equation}
For the following it is useful to express the field amplitude $\mathcal{E}%
_{j}$ in terms of photon flux $\eta _{j}$, Eq. (\ref{eta}), and phase $%
\varphi _{j}$:
\begin{equation*}
\mathcal{E}_{j}=\left| \mathcal{E}_{j}\right| \exp \{-\varphi _{j}\}.
\end{equation*}
Separating the real and imaginary parts, we find from Eq. (\ref{Max3}):
\begin{eqnarray}
\frac{\partial \eta _{j}}{\partial z} &=&-\frac{\partial \mathcal{H}^{\prime
}}{\partial \varphi _{j}},  \label{Can1} \\
\frac{\partial \varphi _{j}}{\partial z} &=&\frac{\partial \mathcal{H}%
^{\prime }}{\partial \eta _{j}}.  \notag
\end{eqnarray}
These equations have the form of Hamilton equations of classical canonical
mechanics with action and angle variables $\eta _{j}$, and $\varphi _{j}$,
''time'' $z$, and the Hamiltonian function $\mathcal{H}^{\prime }=\frac{1}{2}%
N\lambda _{0}$.


\subsection{Constants of motion for resonant 3-wave mixing}


For atomic systems with a closed loop of transitions, the set of canonical
equations (\ref{Can1}) can be further simplified and in fact under some
conditions explicitely integrated. This is the case when the number of
coherent fields involved is not too large. To illustrate the procedure, let
us consider the eigenvalue equation (\ref{Eig1}) for the three-level system
in Fig. 1. In rotating-wave approximation, the light-atom interaction
Hamiltonian is given by:
\begin{eqnarray}
\hat{H} &=&-\hbar \Bigl[\delta _{2}\left| 2\right\rangle \left\langle
2\right| +\delta _{3}\left| 3\right\rangle \left\langle 3\right| \Bigr]
\label{Ham1} \\
&&-\hbar \Omega _{1}\left| 1\right\rangle \left\langle 2\right| +\hbar
\Omega _{2}e^{i\varphi }\left| 2\right\rangle \left\langle 3\right| +\hbar
\Omega _{3}\left| 1\right\rangle \left\langle 3\right| +H.c.,  \notag
\end{eqnarray}
and the eigenvalues are determined by the characteristic equation:
\begin{eqnarray}
&&\lambda _{0}\left( \delta _{2}+\lambda _{0}\right) \left( \delta
_{3}+\lambda _{0}\right) -\left( \Omega _{1}^{2}+\Omega _{2}^{2}+\Omega
_{3}^{2}\right) \lambda _{0}-  \notag \\
&&\quad -\Omega _{1}^{2}\delta _{3}-\Omega _{3}^{2}\delta _{2}=-2\Omega
_{1}\Omega _{2}\Omega _{3}\cos \varphi .  \label{Eig2}
\end{eqnarray}
Here $\Omega _{j}$ are the Rabi frequencies, related to the photon flux via
the coefficients $\mu _{j}$, Eq. (\ref{mus}), as $\Omega _{j}=\sqrt{\mu
_{j}\eta _{j}}$.

The relative phase $\varphi $ of the \textit{em} waves is
\begin{equation}
\varphi =\varphi _{1}+\varphi _{2}-\varphi _{3}-\Delta kz,  \label{phase1}
\end{equation}
which includes the residual phase mismatch $\Delta k$. Also the multiphoton
resonance condition
\begin{equation}
\omega _{3}=\omega _{1}+\omega _{2}  \label{mph}
\end{equation}
has been used.

One can see from Eq. (\ref{Eig2}) that $\lambda _{0}$ and, hence $\mathcal{H}%
^{\prime }$, depend on the field phases $\varphi _{j}$ only through the
relative phase $\varphi $. Therefore, we have:
\begin{equation}
\frac{\partial \mathcal{H}^{\prime }}{\partial \varphi _{1}}=\frac{\partial
\mathcal{H}^{\prime }}{\partial \varphi _{2}}=-\frac{\partial \mathcal{H}%
^{\prime }}{\partial \varphi _{3}}\left( =\frac{\partial \mathcal{H}^{\prime
}}{\partial \varphi }\right) .  \label{phc}
\end{equation}
An immediate consequence of this symmetry of $\mathcal{H}^{\prime }$ is the
existence of constants of motion. Substituting the above equations (\ref{phc}%
) into the first line of Eqs. (\ref{Can1}) yields the well-known Manley-Rowe
relations \cite{boyd}:
\begin{equation}
\frac{\partial \eta _{1}}{\partial z}=\frac{\partial \eta _{2}}{\partial z}=-%
\frac{\partial \eta _{3}}{\partial z},  \label{MR}
\end{equation}
which correspond to two independent constants of motion:
\begin{eqnarray}
\eta _{1}+\eta _{3} &=&\eta _{10}+\eta _{30},  \label{const1} \\
\eta _{1}-\eta _{2} &=&\eta _{10}-\eta _{20}.  \notag
\end{eqnarray}
Here $\eta _{j0}=\eta _{j}(z=0)$ are the photon flux values at the entrance
to the medium. Taking into account the multiphoton resonance condition (\ref
{mph}), one finds furthermore that the total intensity of the \textit{em}
fields is conserved: $I_{1}+I_{2}+I_{3}=const(z)$. The Manley-Rowe relations
and the constants of motion tell us that in the process under consideration,
the energy is transferred from the frequency components $\omega _{1}$, $%
\omega _{2}$ into $\omega _{3}$ and back, with equal rates and without
losses, which of course is expected for a dissipationless nonlinear medium.

The relations Eq. (\ref{const1}) enable us to re-write $\eta _{j}$ as:
\begin{eqnarray}
\eta _{1}(z) &=&\eta _{10}-J(z),  \notag \\
\eta _{2}(z) &=&\eta _{20}-J(z),  \label{etas} \\
\eta _{3}(z) &=&\eta _{30}+J(z).  \notag
\end{eqnarray}
$J(z)$ characterizes the amount of energy exchange between the waves and has
the initial condition $J(z=0)=0$.

Thus the original problem with six amplitude and phase variables can be
reduced to two variables $J$ and $\varphi $ by a canonical transformation.
This leads to
\begin{eqnarray}
\frac{\partial J}{\partial z} &=&-\frac{\partial \mathcal{H}}{\partial
\varphi },  \label{Can2J} \\
\frac{\partial \varphi }{\partial z} &=&\frac{\partial \mathcal{H}}{\partial
J},  \label{Can2phi}
\end{eqnarray}
with the new Hamiltonian function
\begin{equation}
\mathcal{H}=\frac{1}{2}N\lambda _{0}+\Delta kJ\equiv \frac{1}{2}N\lambda .
\label{Ham2}
\end{equation}
As can be seen from Eqs. (\ref{Ham2}) and (\ref{Eig2}), $\mathcal{H}$ (or $%
\lambda $) does not depend on the coordinate $z$ explicitly. Therefore, $%
\mathcal{H}$ (or $\lambda $) is a fourth constant of motion expressing the
conservation of the energy density of the medium with respect to $z$.


\subsection{Solution of the wave-propagation problem}


To solve the remaining two equations of motion for $J(z)$ and $\varphi (z)$,
the Rabi-frequencies $\Omega _{j}$ are expressed in terms of $\eta _{j0}$
and $J$, and the characteristic equation (\ref{Eig2}) is written in the form
\begin{equation}
G(\lambda ,J)=g(J)\cos \varphi .  \label{Eig3}
\end{equation}
Differentiating both sides with respect to $\varphi $ yields
\begin{equation*}
\frac{\partial G}{\partial \varphi }=\frac{\partial G}{\partial \lambda }%
\frac{\partial \lambda }{\partial \varphi }=-g\sin \varphi =\pm \sqrt{%
g^{2}-G^{2}}.
\end{equation*}
Substituting this relation into Eq. (\ref{Can2J}), we find:
\begin{equation}
\frac{\partial J}{\partial z}=\pm \frac{N}{2}\frac{\sqrt{g^{2}-G^{2}}}{%
\partial G/\partial \lambda }.  \label{fin}
\end{equation}
The choice of the sign in Eq. (\ref{fin}) depends on the sign of $\sin
\varphi $ at $z=0$. Integration of Eq. (\ref{fin}) gives an implicit
solution for $J(z)$:
\begin{equation}
\pm \frac{N}{2}z=\int\limits_{0}^{J}\frac{\partial G(J^{\prime })}{\partial
\lambda }\frac{dJ^{\prime }}{\sqrt{g^{2}\left( J^{\prime }\right)
-G^{2}\left( J^{\prime }\right) }}.  \label{int}
\end{equation}
To analytically evaluate the remaining integral, we note that both functions
$g^{2}-G^{2}$ and $\partial G/\partial \lambda $ are polynomials in $J$:
\begin{eqnarray*}
g &=&-2\sqrt{\mu _{1}\mu _{2}\mu _{3}}\sqrt{\left( \eta _{10}-J\right)
\left( \eta _{20}-J\right) \left( \eta _{30}+J\right) }, \\
G &=&G_{0}+\sum_{m=1}^{3}A_{m}J^{m}, \\
\frac{\partial G}{\partial \lambda } &=&\sum_{m=0}^{2}a_{m}J^{m}.
\end{eqnarray*}
Therefore, equation (\ref{fin}) describes a one-dimensional finite motion of
a pendulum in an external potential. The solution is in general given by
some combination of elliptic functions \cite{ell} with parameters determined
mainly by the roots $J_{n}$ of the polynomial equation:
\begin{equation}
g^{2}\left( J\right) -G^{2}\left( J\right) =0.  \label{root1}
\end{equation}
The allowed range of $J$, corresponding to the region of classically allowed
motion of the pendulum, lies between $zero$ and the smallest positive root $%
J_{1}$ of the polynomial (\ref{root1}).

The eigenvalue $\lambda $ is a constant of motion (cf. Eq. (\ref{Ham2})),
and can thus be found from the characteristic equation (\ref{Eig3}) with
parameters taken at the medium entrance $z=0$:
\begin{equation}
G_{0}\left( \lambda \right) =g(z=0)\cos \varphi (z=0).  \label{lamb}
\end{equation}

Thus, we have reduced the propagation problem to solving two algebraic
equations: (\ref{lamb}) for $\lambda $ and (\ref{root1}) for the roots $J_{n}
$. If this can be done explicitly, the Hamiltonian method provides an
analytical solution to the propagation problem. But even if an explicit
solution is not possible, it considerably simplifies numerical calculations.
Apart from the advantage of being convenient from the formal point of view,
the Hamiltonian method allows a deeper insight into the nature of the wave
propagation process. For example, the method provides a direct access to the
stability analysis of the solution by referring to a well-developed theory
of Hamiltonian systems \cite{cap91}.

To understand the dynamics of the system, in particular interesting
quantities such as the conversion efficiency, it is useful to discuss the
physical meaning of the coefficients $A_{m}$ and $a_{m}$. Considering the
canonical equation (\ref{Can2phi}) for the relative phase:
\begin{equation}
\frac{\partial \varphi }{\partial z}=\frac{N}{2}\frac{\partial \lambda }{%
\partial J}=\frac{N}{2}\frac{\partial G/\partial J}{\partial G/\partial
\lambda }=\frac{N}{2}\frac{A_{1}+2A_{2}J+3A_{3}J^{2}}{a_{0}+a_{1}J+a_{2}J^{2}%
},  \label{refr}
\end{equation}
one recognizes that the $A_{m}$ and $a_{m}$ describe the linear and
nonlinear refraction coefficients of the medium. E.g. if $J$ is sufficiently
small, the first term $NA_{1}/2a_{0}$ on the right-hand side of Eq. (\ref
{refr}) can be identified with the phase mismatch induced by the linear
refraction, including both contributions from the three-level interaction
and the residual mismatch $\Delta k$.


\subsection{Generation of field with frequency $\protect\omega _{3}$}


In the context of the present discussion we are most interested in the
generation of the $\omega _{3}$\ mode from vacuum $\eta _{30}=0$. In this
case, the eigenvalue equation (\ref{lamb}) reduces to
\begin{eqnarray}
G_{0}(\lambda ) &=&\lambda \left( \lambda +\delta _{2}\right) \left( \lambda
+\delta _{3}\right) -\mu _{1}\eta _{10}\left( \lambda +\delta _{3}\right)
\notag \\
&&-\mu _{2}\eta _{20}\lambda =0.  \label{F0}
\end{eqnarray}
The nonvanishing coefficients in the expansion of $G$ and $\partial
G/\partial \lambda $ are given by
\begin{eqnarray}
A_{1} &=&q\Bigl[\left( \lambda +\delta _{2}\right) \left( \lambda +\delta
_{3}\right) +\lambda \left( \lambda +\delta _{2}\right) +\lambda \left(
\lambda +\delta _{3}\right)  \notag \\
&&\qquad -\mu _{1}\eta _{10}-\mu _{2}\eta _{20}\Bigr]  \label{A1} \\
&&+\mu _{1}\left( \lambda +\delta _{3}\right) +\mu _{2}\lambda -\mu
_{3}\left( \lambda +\delta _{2}\right) ,  \notag \\
A_{2} &=&q^{2}\left( 3\lambda +\delta _{2}+\delta _{3}\right) +q\left( \mu
_{1}+\mu _{2}-\mu _{3}\right) ,  \label{A2} \\
A_{3} &=&q^{3},  \label{A3} \\
a_{0} &=&3\lambda ^{2}+2\lambda \left( \delta _{2}+\delta _{3}\right)
+\delta _{2}\delta _{3}-\mu _{1}\eta _{10}-\mu _{2}\eta _{20},  \label{a0} \\
a_{1} &=&2q\left( 3\lambda +\delta _{2}+\delta _{3}\right) +\left( \mu
_{1}+\mu _{2}-\mu _{3}\right) ,  \label{a1} \\
a_{2} &=&3q^{2},  \label{a2}
\end{eqnarray}
where
\begin{equation}
q=-2\Delta k/N.  \label{q}
\end{equation}

With this one finds for the denominator in the integral (\ref{int}), which
determines the allowed values of $J$:
\begin{eqnarray}
g^{2}-G^{2} &=&4\mu _{1}\mu _{2}\mu _{3}J\left( \eta _{10}-J\right) \left(
\eta _{20}-J\right)  \notag \\
&&-\left( A_{1}+A_{2}J+A_{3}J^{2}\right) ^{2}J^{2}=0.  \label{root2}
\end{eqnarray}
The second term in this expression is non-positive, so the smallest positive
root of the polynomial is bounded by the minimum of $\eta _{10}$ and $\eta
_{20}$. This reflects the fact that the conversion process stops when the
energy of the weaker of the two pump fields is entirely depleted. In order
to reach this limit and thus in order to attain maximum conversion
efficiency, the second term in (\ref{root2}) should be small, which
corresponds to the phase mismatch to be negligible:
\begin{equation}
A_{1}+A_{2}J+A_{3}J^{2}\approx 0  \label{comp}
\end{equation}
In order to see what parameters are required to approximately satisfy this
condition, we have to analyze the coefficients $A_{m}$.

It is worthwhile to point out that the behavior of the total phase is
described by Eq. (\ref{Eig3}):
\begin{equation}
\cos \varphi \left( z\right) =\frac{G}{g}=\frac{A_{1}J+A_{2}J^{2}+A_{3}J^{3}%
}{2\sqrt{\mu _{1}\mu _{2}\mu _{3}J\left( \eta _{10}-J\right) \left( \eta
_{20}-J\right) }}.  \label{cos-phi}
\end{equation}
When the total phase mismatch is compensated, $A_{1}+A_{2}J+A_{3}J^{2}%
\approx 0$, the phase follows the equation:
\begin{equation*}
\cos \varphi \left( z\right) =0,
\end{equation*}
that is, the phase is constant and equal to $+\pi /2$ or $-\pi /2$ when all
of the waves are present, and jumps by $\pi $ to $-\pi /2$ or $+\pi /2$ when
one of the wave intensities approaches zero.

As it is seen from Eqs. (\ref{A2}), (\ref{A3}), the terms $A_{2}$, $A_{3}$
responsible for the intensity-dependent refractive index are both
proportional to the residual phase mismatch $\Delta k$. One can show that in
atomic (molecular) media, where the background refractive index is given by
Eq. (\ref{n}), $\Delta k$ is always sufficiently small so that the leading
term in the phase mismatch is $A_{1}$, while $A_{2}J$ and $A_{3}J^{2}$ are
negligibly small. It is also true, that the influence of the
intensity-dependent phase mismatch remains insignificant even when the
linear refraction is compensated: $A_{1}=0$. Therefore, in what follows, we
will disregard the terms $A_{2}J$ and $A_{3}J^{2}$. In this case, the
solution of the propagation equation (\ref{int}) gives the following
dependence of $J(z)$ in implicit form:
\begin{eqnarray}
&&\pm \kappa z+\chi _{0}=F\left[ \gamma _{1}\left( J\right) ,p\right] +\frac{%
a_{1}J_{2}}{a_{0}}\left\{ F\left[ \gamma _{2}\left( J\right) ,p\right]
\right.  \notag \\
&&\quad \left. +\left( 1-\frac{J_{1}}{J_{2}}\right) \Pi \left[ \gamma
_{2}\left( J\right) ,p^{2},p\right] \right\} ,  \label{sol00}
\end{eqnarray}
where $\chi _{0}$ is an integration constant, and $F\left( \gamma ,p\right) $
and $\Pi \left( \gamma ,r,p\right) $ are the elliptic integrals of the first
and third kind, respectively \cite{ell}. $\kappa $ is the nonlinear
conversion coefficient defined as
\begin{equation}
\kappa =\frac{N}{2}\frac{\sqrt{\mu _{1}\mu _{2}\mu _{3}J_{2}}}{a_{0}},
\label{kappa}
\end{equation}
The parameters of the elliptic integrals are:
\begin{eqnarray}
\gamma _{1}\left( J\right) &=&\arcsin \sqrt{\frac{J}{J_{1}}},\;  \notag \\
\gamma _{2}\left( J\right) &=&\arcsin \sqrt{\frac{J_{2}\left( J_{1}-J\right)
}{J_{1}\left( J_{2}-J\right) }},\quad  \label{par1} \\
p &=&\sqrt{\frac{J_{1}}{J_{2}}}.  \notag
\end{eqnarray}
with $J_{1}$ and $J_{2}$ being the roots of Eq. (\ref{root2}):
\begin{eqnarray}
J_{2,1} &=&\frac{1}{2}\left( \eta _{10}+\eta _{20}+B_{1}\right)  \notag \\
&&\pm \frac{1}{2}\sqrt{\left( \eta _{10}+\eta _{20}+B_{1}\right) ^{2}-4\eta
_{10}\eta _{20}},  \label{J21} \\
B_{1} &=&\frac{A_{1}^{2}}{4\mu _{1}\mu _{2}\mu _{3}}.  \notag
\end{eqnarray}

In the next section we discuss the solution of the propagation problem for
several field-atom interaction configurations resulting in the EIT-assisted,
maximum coherence, and conventional nonlinear-optical regimes of the
frequency conversion.


\section{Analytical solutions for resonant 3-wave mixing}


We assume in our simple 3-level model system that the transitions $\left|
1\right\rangle -\left| 3\right\rangle $, $\left| 2\right\rangle -\left|
3\right\rangle $ as well as $\left| 1\right\rangle -\left| 2\right\rangle $
are allowed. To take into account the much weaker coupling between levels $%
|1\rangle $ and $|2\rangle $ which in reality is a two-photon transition, we
assume $\mu _{1}\ll \mu _{2}$. An analytic solution for the more general
4-wave mixing system can also be obtained within the Hamiltonian formalism
\cite{to-be-published}. The explicit results are however rather lengthy and
not very instructive.


\subsection{EIT-based upconversion}


The EIT scheme implies two-photon resonance between the states $\left|
1\right\rangle $ and $\left| 3\right\rangle $: $\delta _{3}=0$. According to
the idea of EIT-assisted nonlinear optics, we suppose here that $\eta
_{20}\geq \eta _{10}$. Since $\mu _{2}\gg $ $\mu _{1}$, we always have $%
\Omega _{2}\gg \Omega _{1}$. We allow, for the moment, for
spontaneous decay from state $\left| 3\right\rangle $ out of the
system, but require the initial Rabi frequency $\Omega _{20}$ to
be much larger than the decay rate: $\Omega _{20}\gg \gamma $.
Under these conditions, the eigenstate that asymptotically
connects to the ground state of atoms for $t\rightarrow -\infty $
is the state
\begin{equation*}
\left| \psi _{0}(z=0)\right\rangle \approx \frac{\Omega _{20}}{\sqrt{\Omega
_{10}^{2}+\Omega _{20}^{2}}}\left| 1\right\rangle -\frac{\Omega
_{10}e^{-i\varphi }}{\sqrt{\Omega _{10}^{2}+\Omega _{20}^{2}}}\left|
3\right\rangle .
\end{equation*}
corresponding to the complex eigenvalue
\begin{equation*}
\lambda \approx i\gamma \frac{\Omega _{10}^{2}}{\Omega _{20}^{2}}.
\end{equation*}
Since $\Omega _{10}\ll \Omega _{20}$ one has: $\left| \lambda
\right| \ll \gamma $. Correspondingly, the probability of
spontaneous decay of the state $\left| \psi _{0}\right\rangle $,
which can be defined as Im($\lambda $)$\tau $ with $\tau $ being
the characteristic pulse length, remains small even for rather
long pulses: $\tau \ll \tau _{0}\equiv \left( \Omega
_{20}^{2}/\Omega _{10}^{2}\right) \gamma ^{-1}$ with $\tau _{0}\gg
\gamma ^{-1}$. For such pulses, we can safely disregard the
spontaneous emission and put $\lambda \approx 0$, which means that
the corresponding adiabatic state is the usual dark state of EIT.
Moreover, since $\,\Omega _{10}\ll $ $\Omega _{20}$ the eigenstate
$\left| \psi _{0}(z=0)\right\rangle $ is an only slightly
disturbed ground state $\left| 1\right\rangle $. Therefore, the
requirement of adiabaticity of the process is satisfied
automatically at the medium entrance, $z=0$.

The coefficients $A_{1}$ and $a_{m}$ are given for the regime of EIT by the
following expressions:
\begin{eqnarray}
A_{1} &\simeq &-q\mu _{2}\eta _{20}-\mu _{3}\delta _{2},  \label{A1e} \\
a_{0} &\simeq &-\mu _{2}\eta _{20},  \label{a0e} \\
a_{1} &=&2q\delta _{2}+\left( \mu _{1}+\mu _{2}-\mu _{3}\right) .
\label{a1e}
\end{eqnarray}

\subsubsection{Undepleted EIT-generating field}

In the original proposal of EIT-assisted frequency conversion, the $\omega
_{2}$ field is assumed to be very strong and undepleted. The latter
condition implies $\eta _{10}\ll \eta _{20}$. In this situation the solution
(\ref{sol00}) can be well approximated by
\begin{eqnarray}
J &=&\frac{\eta _{10}}{1+\left( \frac{\Delta k^{\prime }}{2\kappa _{e}}%
\right) ^{2}}\sin ^{2}\left( \kappa _{e}z\sqrt{1+\left( \frac{\Delta
k^{\prime }}{2\kappa _{e}}\right) ^{2}}\right) ,  \label{sole1} \\
\kappa _{e} &=&\frac{N}{2}\sqrt{\frac{\mu _{1}\mu _{3}}{\mu _{2}\eta _{20}}},
\label{kape} \\
\Delta k^{\prime } &=&\Delta k-\frac{N}{2}\frac{\mu _{3}\delta _{2}}{\mu
_{2}\eta _{20}},  \label{dke}
\end{eqnarray}
which, of course, coincides with that obtained in Sect. II, Eqs. (\ref{sol1}%
) and (\ref{kap-e}), under the undepleted drive approximation. The influence
of the intensity-dependent phase mismatch (terms $A_{2}J$ and $A_{3}J^{2}$)
is negligible in this case, as was discussed above.

One can immediately see from Eqs. (\ref{A1e}), (\ref{A2}), (\ref{A3}) that
for vanishing residual phase mismatch $q=0$ (which can be realized by adding
a buffer gas with proper dispersion \cite{buf}) and for $\delta _{2}=0$
there is perfect phase matching: $A_{1}=A_{2}=A_{3}=0$. Consequently maximum
energy transfer into the generated wave $\omega _{3}$ is possible. This
applies also for the case when linear refraction is compensated by a
detuning $\delta _{2}$:
\begin{equation}
\delta _{2}=\frac{2}{N}\frac{\mu _{2}\eta _{20}}{\mu _{3}}\Delta k.
\label{d2}
\end{equation}
For both of these phase matching procedures, the maximum value of $\eta
_{3}=J$ achieved at $z_{e}=\pi /2\kappa _{e}$ is equal to $\eta _{10}$. Thus
the instantaneous fractional conversion efficiency, defined as
\begin{equation}
\epsilon \equiv \frac{J_{\mathrm{max}}}{\min \left( {\eta _{10},\eta _{20}}%
\right) }  \label{frace}
\end{equation}
can become unity. It should be noted, that $\epsilon =1$ is in general only
achieved at a specific instant of time since $\kappa _{e}$ depends on time
via $\eta _{20}\left( t-z/c\right) \equiv \bar{\eta}_{20}\,f_{20}\left(
t-z/c\right) $ ($\bar{\eta}_{20}$ is the amplitude and $f_{20}\left(
t-z/c\right) $ is the temporal envelope of the $\eta _{20}\left(
t-z/c\right) $ pulse). Hence the transfer of energy is complete only for the
part of the $\omega _{1}$ pulse which satisfies the condition $\sqrt{%
f_{20}\left( t-z/c\right) }\approx 1$. A high overall conversion requires a
flat temporal profile of the ''coupling'' pulse $\omega _{2}$. This is easy
to implement, however, since always $\eta _{3}=J\ll \eta _{2}$ ($\approx
\eta _{20}$) and $\eta _{1}\ll \eta _{20}$. We have then $\Omega _{1},\Omega
_{3}\ll \Omega _{2}$ for the whole medium which means, in particular, that
all the atoms throughout the medium are essentially in the ground state.
Therefore, the $\omega _{2}$ pulse can be arbitrarily long.

In contrast to the fractional conversion efficiency, Eq. (\ref{frace}), the
total conversion efficiency, defined as
\begin{equation}
W(z)\equiv \frac{\displaystyle{\int \mathrm{d}t\,\omega _{3}\eta _{3}(z,t)}}{%
\displaystyle{\ \int \mathrm{d}t\,\Bigl[\omega _{1}\eta _{1}(z,t)+\omega
_{2}\eta _{2}(z,t)\Bigr]}}  \label{W}
\end{equation}
remains very small because $\eta _{10}\ll \eta _{20}$.

\subsubsection{Depletion of the pump fields}

\begin{figure}[th]
\centerline{\epsfig{file=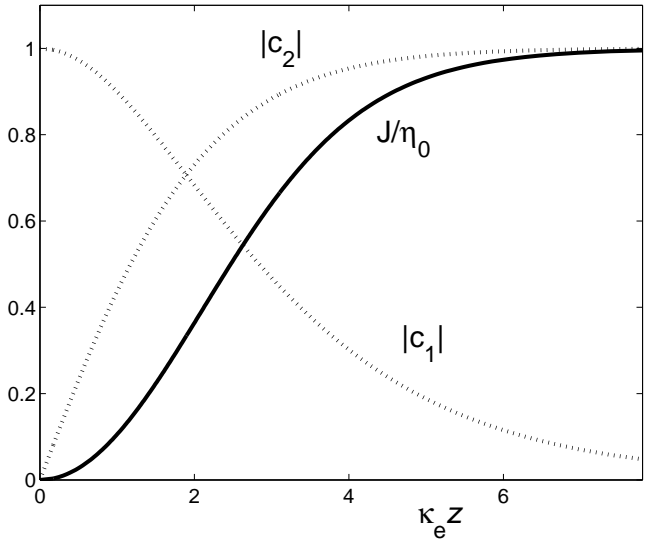,width=8.5 cm}} \vspace*{2ex}
\caption{EIT-upconversion for perfect phase matching. Solid line: spatial
evolution of $J$ according to Eq. (\ref{sole2}) for given retarded time. For
sufficiently large propagation length there is complete conversion. Dotted
lines: corresponding spatial evolution of the atomic bare state
probabilities. The amplitude of the excited state $\left| 3\right\rangle $
is negligibly small (not shown). Parameters: $\protect\mu _{2}/\protect\mu
_{3}=0.5$, $\protect\mu _{1}/\protect\mu _{3}=0.05$.}
\label{conv1}
\end{figure}

Conversion with $\varepsilon \approx 1$ can be achieved, however, also for
the case when both pump fields $\omega _{1}$ and $\omega _{2}$ are of
comparable intensity. For example, for exact phase matching and $\eta
_{10}=\eta _{20}\equiv $ $\eta _{0}$ (which would correspond, e.g., to the
second harmonic generation), the solution Eq. (\ref{sol00}) can be written
as:
\begin{equation}
\frac{\mu _{2}}{\mu _{3}}\kappa _{e}z=\text{Arth}\left( \sqrt{\frac{J}{\eta
_{0}}}\right) -\frac{\left( \mu _{1}+\mu _{2}-\mu _{3}\right) }{\mu _{3}}%
\sqrt{\frac{J}{\eta _{0}}},  \label{sole2}
\end{equation}
which demonstrates that $J$ monotonically approaches $\eta _{0}$ as $z$
increases. The form of this solution for given retarded time is shown in
Fig. \ref{conv1}. The monotonic dependence ensures that after sufficiently
long propagation length all parts of the pulse $\eta _{10}=\eta _{20}$ are
converted into the $\omega _{3}$ wave so that not only $\varepsilon \approx
1 $, but also the total energy conversion efficiency $W$ reaches unity.


\begin{figure}[th]
\centerline{\epsfig{file=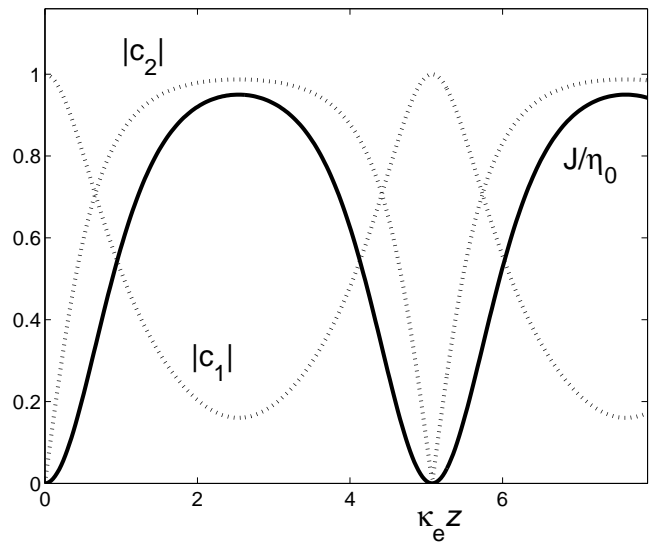,width=8.5 cm}} \vspace*{2ex}
\caption{EIT-upconversion with finite phase mismatch. Solid line: spatial
evolution of $J$. Dotted lines: corresponding spatial evolution of the
atomic bare state probabilities. The amplitude of the excited state $\left|
3\right\rangle $ is negligibly small (not shown). Parameters are the same as
in Fig. \ref{conv1}, $\Delta k^{\prime }/2\protect\kappa _{e}=0.06$. }
\label{Jasn}
\end{figure}

When the phase mismatch is not compensated, the solution for $\eta
_{10}=\eta _{20}\equiv $ $\eta _{0}$ is given by more general expression Eq.
(\ref{sol00}) with $B_{1}=\eta _{0}\left( \Delta k^{\prime }/2\kappa
_{e}\right) ^{2}$ and $J_{2,1}\approx \eta _{0}\left( 1\pm \Delta k^{\prime
}/2\kappa _{e}\right) $. The spatial dependence of the generated field
intensity (Fig. \ref{Jasn}) differs only slightly from
\begin{equation*}
J=\eta _{0}\left( 1-\frac{\Delta k^{\prime }}{2\kappa _{e}}\right) \text{sn}%
^{2}\left[ \kappa _{e}z\sqrt{1+\frac{\Delta k^{\prime }}{2\kappa _{e}}};%
\sqrt{\frac{J_{1}}{J_{2}}}\right] ,
\end{equation*}
where sn$\left[ x;p\right] $ is the Jacobi elliptic sine function. The
period of intensity oscillations seen in Fig. \ref{Jasn} can be estimated as
\cite{ell}
\begin{equation*}
z_{e}\simeq \ln \left( 16\kappa _{e}/\Delta k^{\prime }\right) .
\end{equation*}
Since in EIT regime we have always: $\Delta k^{\prime }\ll 2\kappa _{e}$,
very high total conversion efficiency is achieved without phase matching
also for $\eta _{10}\approx \eta _{20}$ when the drive fields are
substantially depleted.


\subsubsection{Evolution of the atomic state}

In the case of EIT with comparable intensities of the $\omega _{1}$ and $%
\omega _{2}$ pump fields, the intensity of the $\omega _{3}$ wave increases
with $z$ and eventually exceeds the one of the $\omega _{1},\omega _{2}$
waves. Therefore, the adiabatic state of the atomic system will no longer
approximately coincide with the ground state $\left| 1\right\rangle $ over
the entire medium length at a given retarded time. It will evolve via
different superpositions of the bare states $\left| 1\right\rangle $ and $%
\left| 2\right\rangle $ following the change of the fields (Fig.
\ref{conv1} and \ref{Jasn}) \cite{kors}. We see that the
''conventional'' EIT regime (one strong and one weak field, and
thus one highly populated and one slightly populated state) does
not hold over the full conversion cycle. We should stress,
however, that the atoms remain in the adiabatic state
corresponding to $\lambda =0$, i.e., in the dark state. Therefore,
all the features of the EIT-assisted conversion process such as
vanishing linear absorption and refraction as well as resonantly
enhanced nonlinearity are present even when the pump field is
depleted.

When most of the input energy is transferred to the $\omega _{3}$ pulse,
i.e. for larger propagation distances into the medium, the dark state $|\psi
_{0}\rangle $ approaches the bare state $|2\rangle $
\begin{equation*}
|\psi _{0}\rangle \,\rightarrow \,|2\rangle ,
\end{equation*}
which does not coincide with the initial atomic state (i.e. the ground state
$|1\rangle $). Thus we encounter two problems: First, since all atoms are in
the ground state before the arrival of the pulses, there must be an
adiabatic transfer in the initial phase from $|1\rangle $ to $|\psi
_{0}\rangle $. This requires a specific time-order of the pulses. In
particular $\Omega _{2}$ should arrive after the other pulses. This time
ordering is however not reflected in our solutions. Second, after the
interaction with the pulses, there is energy left in the atomic system. This
is in apparent contradiction to the fact that full conversion of photon
energy has taken place and no energy got lost. It should be noted though
that the size of the effect is rather small in the considered limit of a
large initial number of photons. This contradictory behavior is a result of
the limited validity of the assumptions made. We have here completely
neglected non-adiabatic corrections, which are small on the level of
individual atoms, but add up in the field evolution when integrated over the
whole sample. The most important effect ignored by this is the group
velocity delay of the weak fields.

Although the group delay cannot be treated within the current approach, it
may resolve the problems. In the first part of the medium, i.e. for small $%
Nz $ the generated field is still small. Thus the group velocities of the
pulses $\Omega _{1}$ and $\Omega _{3}$ should be smaller than that of the
strong drive field $\Omega _{2}$. Then the leading edge of the $\Omega _{2}$%
-pulse will always propagate ahead of the other pulses, guaranteeing that
the adiabatic state $|\psi _{0}\rangle $ asymptotically matches with the
ground state at early times for all values of $z$. For larger propagation
distances into the medium and for times in the central part of the pulses a
substantial portion of the energy is transferred to the field $\Omega _{3}$.
Now the situation is reversed. $\Omega _{3}$ is strong and $\Omega _{2}$ is
weak. Thus the pulse $\Omega _{2}$ will have a smaller group velocity and
should lag behind at the trailing edge of the pulses. Consequently also for
large times the adiabatic eigenstate $|\psi _{0}\rangle $ approaches
asymptotically the ground state $|1\rangle $.

A full quantitative account of the leading order non-adiabatic corrections
requires however a reformulation of the Hamiltonian approach. This is beyond
the scope of the present work and will be presented in a future publication.


\subsection{Maximum coherence case}

The mechanism of the nonlinear optics with maximum coherence is realized
when the detuning $\delta _{3}$ is much larger than the other parameters
including all Rabi frequencies and $\delta _{2}$. In this case the
coefficients $A_{1}$ and $a_{m}$ are as follows:
\begin{eqnarray}
A_{1} &\simeq &q\left[ \left( 2\lambda +\delta _{2}\right) \delta _{3}-\mu
_{1}\eta _{10}-\mu _{2}\eta _{20}\right]  \label{A1m} \\
&&+\mu _{1}\delta _{3}+\mu _{2}\lambda -\mu _{3}\left( \lambda +\delta
_{2}\right) ,  \notag \\
a_{0} &\simeq &\left( 2\lambda +\delta _{2}\right) \delta _{3}-\mu _{1}\eta
_{10}-\mu _{2}\eta _{20},  \label{a0m} \\
a_{1} &=&2q\delta _{3}+\left( \mu _{1}+\mu _{2}-\mu _{3}\right) .
\label{a1m}
\end{eqnarray}

\subsubsection{Preparation of maximum atomic coherence}

The energy eigenvalue for large $\delta _{3}$ is given by:
\begin{equation}
\lambda =-\frac{1}{2}\left( \delta _{2}-\frac{\Omega _{20}^{2}}{\tilde{\delta%
}_{3}}\right) +\frac{1}{2}\sqrt{\left( \delta _{2}-\frac{\Omega _{20}^{2}}{%
\tilde{\delta}_{3}}\right) ^{2}+4\Omega _{10}^{2}},  \label{lam1}
\end{equation}
where $\tilde{\delta}_{3}=\delta _{3}+i\gamma $. The amplitudes of the
adiabatic state corresponding to this eigenvalue at the entrance to the
medium, $z=0,$ are:
\begin{eqnarray}
c_{1} &=&\frac{\Omega _{10}}{\sqrt{\lambda ^{2}+\Omega _{10}^{2}}},
\label{c1m} \\
c_{2} &=&-\frac{\lambda }{\sqrt{\lambda ^{2}+\Omega _{10}^{2}}},  \label{c2m}
\\
c_{3} &\ll &c_{1},c_{2}.  \notag
\end{eqnarray}
Similar to the EIT case, the probability of spontaneous decay of the state $%
\left| \psi _{0}\right\rangle $ is small: Im($\lambda $)$\tau \ll
1$ for large detuning assumed here, $\delta _{3}\gg \gamma $, even
for pulses that are longer than $\gamma ^{-1}$: $\tau \ll \left(
\delta _{3}^{2}/\Omega _{20}^{2}\right) \gamma ^{-1}$. Again, we
will consider only such pulses and neglect spontaneous emission.

The situation of the maximum coherence, $\left| c_{1}\right|
=\left| c_{2}\right| =1/\sqrt{2}$, takes place when the Rabi
frequency exceeds the detuning on transition $\left|
1\right\rangle -\left| 2\right\rangle $ including the ac Stark
shift induced by the $\omega _{2}$ field: $\Omega _{10}\gg \left|
\delta _{2}-\Omega _{20}^{2}/\delta _{3}\right| $. In this case
$\lambda \approx \Omega _{10}$ and the atomic state at the medium
entrance is the superposition: $\left| \psi _{0}(z=0)\right\rangle
=\left( \left| 1\right\rangle -\left| 2\right\rangle \right)
/\sqrt{2}$. However, in order for the atoms to get there from the
ground state, one first has to have the inverse situation $\delta
_{2}\gg \Omega _{10}$, and then adiabatically decrease $\delta
_{2}$, simultaneously increasing $\Omega _{10} $. This can be
accomplished by, e.g., Stark-chirped rapid adiabatic passage
\cite{scrap}. We will not include this process explicitly in the
consideration, but assume that when the $\omega _{2}$ pulse
arrives, the condition $\Omega _{10}\gg \delta _{2}$ is already
satisfied. As a consequence, the energy taken from the leading
edge of the $\omega _{1}$ pulse for the superposition preparation,
is not taken into account. However, if the number of photons in
the $\omega _{1}$ pulse is much larger than the total number of
atoms along the field propagation path, these preparation energy
losses ($\hbar \omega _{1}/2$ per atom) can be neglected.

\subsubsection{Undepleted coherence-generating field}

When a strong, undepleted $\omega _{1}$ field is supposed: $\eta _{20}\ll
\eta _{10}$ (implying also undepleted coherence on $\left| 1\right\rangle
-\left| 2\right\rangle $ transition), the generated field intensity
experiences sinusoidal oscillations with respect to the propagation length:
\begin{eqnarray}
J &=&\frac{\eta _{20}}{1+\left( \frac{\Delta k^{\prime }}{2\kappa _{m}}%
\right) ^{2}}\sin ^{2}\left( \kappa _{m}z\sqrt{1+\left( \frac{\Delta
k^{\prime }}{2\kappa _{m}}\right) ^{2}}\right) ,  \label{solm1} \\
\kappa _{m} &=&\frac{N}{4}\frac{\sqrt{\mu _{2}\mu _{3}}}{\delta _{3}},
\label{kapm} \\
\Delta k^{\prime } &=&\Delta k-\frac{N}{4}\left( \sqrt{\frac{\mu _{1}}{\eta
_{10}}}+\frac{\left( \mu _{2}-\mu _{3}\right) }{\delta _{3}}\right) ,
\label{dkm}
\end{eqnarray}
in correspondence with the analysis of Sec. II, Eqs. (\ref{sol1}), (\ref
{kap-m}) and (\ref{dk-m}), and assuming $\left| c_{1}\right| ^{2}=\left|
c_{2}\right| ^{2}=\rho _{12}=1/2$.

Since $\kappa _{m}$ is of the order of $\Delta k^{\prime }$, substantial
transfer of energy from the $\omega _{2}$ field into the $\omega _{3}$ field
occurs even without compensation of the phase mismatch. If one can manage to
make $\Delta k^{\prime }$ small: $\Delta k^{\prime }\ll \kappa _{m}$ then
the best possible regime for conversion can be attained: not only complete
transfer for the pulse maximum is realized: $J_{1}\approx J_{max}=\eta _{20}$%
, but also the conversion length $l=\kappa _{m}^{-1}$ does not depend on
time, and hence, all parts of the $\omega _{2}$ pulse are homogeneously
converted into the $\omega _{3}$ wave.

There are several possibilities to reach this regime. First, one may set $%
\Delta k$ to zero by use of the buffer gas, and choose the input laser
parameters such that:
\begin{equation}
\delta _{3}=-\frac{\left( \mu _{2}-\mu _{3}\right) }{\mu _{1}}\sqrt{\mu
_{1}\eta _{10}}.  \label{d3m}
\end{equation}
This condition can be satisfied only for the time interval when $\sqrt{%
f_{10}\left( t-z/c\right) }\approx 1$. That is, a flat temporal profile of
the $\omega _{1}$ pulse is required. Since this is not always possible,
another method can be used. One may simply increase the intensity $\eta
_{10} $ so that the refraction contribution of the $\left| 1\right\rangle
-\left| 2\right\rangle $ transition becomes small: $\sqrt{\mu _{1}/\eta _{10}%
}\ll \left( \mu _{2}-\mu _{3}\right) /\delta _{3}$, and choose the detuning $%
\delta _{3}$ as
\begin{equation}
\delta _{3}=\frac{\left( \mu _{2}-\mu _{3}\right) }{2q}.  \label{qm}
\end{equation}

Since in this regime $\eta _{3}=J\ll \eta _{10}$, we have at any propagation
distance: $\Omega _{1}\gg \Omega _{2}^{2}/\delta _{3}$, $\Omega
_{3}^{2}/\delta _{3}$. Therefore, the atoms remain in a superposition $%
\left| \psi _{0}\right\rangle =\left( \left| 1\right\rangle -\left|
2\right\rangle \right) /\sqrt{2}$ throughout the medium.

It should be noted that although the instantaneous conversion efficiency $%
\epsilon $ can reach unity, the total energy conversion efficiency is always
small, $W\ll 1$, because $J_{1}\approx \eta _{20}\ll \eta _{10}$.

\subsubsection{Depletion of the pump fields}

The situation is completely different for comparable intensities of the two
pump waves: $\eta _{20}\approx \eta _{10}$.

For the maximum coherence regime, $\left| c_{1}\right| =\left| c_{2}\right|
=1/\sqrt{2}$, to be attained, the condition $\Omega _{1}\gg \Omega
_{2}^{2}/\delta _{3}$ must be satisfied (see above), which reduces to $\sqrt{%
\mu _{2}\eta _{10}}\ll \delta _{3}\sqrt{\mu _{1}/\mu _{2}}$ at $\eta
_{20}\approx \eta _{10}$. That is, the intensity of the pump pulses cannot
be large, and the refraction contribution of the $\left| 1\right\rangle
-\left| 2\right\rangle $ transition ($\sim \sqrt{\mu _{1}/\eta _{10}}$) is
not small, so that $\Delta k^{\prime }\approx \left( N/4\right) \sqrt{\mu
_{1}/\eta _{10}}\gg \kappa _{m}$. Thus, the ''exact maximum coherence''
regime fails when the drive field is depleted.

However, the phase matching can still be performed for $\eta _{20}\approx
\eta _{10}$ if the requirement for $\left| c_{1}\right| =\left| c_{2}\right|
=1/\sqrt{2}$ is lifted. One of the possibilities is to cancel the residual
refraction by addition of buffer gas: $\Delta k=0$ (then $A_{2}=A_{3}=0$),
and to choose the interaction parameters such that $A_{1}\approx 0$. The
latter condition can be satisfied for specific relations between the
detuning $\delta _{2}$ and/or $\delta _{3}$, and the input intensities $\eta
_{10},\eta _{20}$. For example, for $\eta _{10}=\eta _{20}\equiv $ $\eta
_{0} $ and when $\mu _{2}<\mu _{3}$, one may choose the detunings $\delta
_{2}=0$ and $\delta _{3}$ such that
\begin{equation}
\mu _{3}\eta _{0}=\frac{\mu _{1}}{\left( \mu _{3}-\mu _{2}\right) }\delta
_{3}^{2}.  \label{cond1}
\end{equation}
Taking into account condition (\ref{cond1}), the eigenenergy $\lambda $\ is
reduced in this regime to:
\begin{equation}
\lambda =\sqrt{\mu _{1}\eta _{0}}\sqrt{\frac{\mu _{3}}{\left( \mu _{3}-\mu
_{2}\right) }}=\Omega _{10}\zeta ,  \label{lam}
\end{equation}
and the corresponding adiabatic state at the entrance to the medium is:
\begin{equation}
\left| \psi _{0}(z=0)\right\rangle =\frac{1}{\sqrt{1+\zeta ^{2}}}\left(
\left| 1\right\rangle -\zeta \left| 2\right\rangle \right) .  \label{psi0}
\end{equation}
Thus, the atoms are prepared not with maximum coherence, $\rho _{12}=1/2$,
but in a specific superposition with $\rho _{12}=\zeta /\left( 1+\zeta
^{2}\right) $ which is also not small.

\begin{figure}[th]
\centerline{\epsfig{file=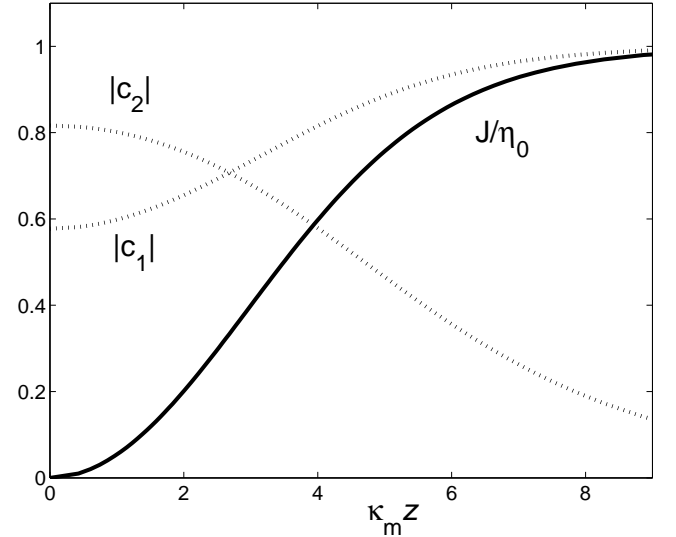,width=8.5 cm}} \vspace*{2ex}
\caption{Upconversion with initial maximum coherence. Solid line: spatial
evolution of $J$ according to Eq. (\ref{solm2}) for given retarded time. For
sufficiently large propagation length there is complete conversion. Dotted
lines: corresponding spatial evolution of the atomic bare state
probabilities. The amplitude of the excited state $\left| 3\right\rangle $
is negligibly small (not shown). Parameters: $\protect\mu _{2}/\protect\mu
_{3}=0.5$, $\protect\mu _{1}/\protect\mu _{3}=0.05$.}
\label{conv-mc}
\end{figure}

The propagation problem solution in this case resembles that of the EIT
regime, Eq. (\ref{sole2}):
\begin{eqnarray}
\kappa _{m}\frac{2}{\zeta }z &=&\text{Arth}\left( \sqrt{\frac{J}{\eta _{0}}}%
\right) -\frac{\mu _{1}+\mu _{2}-\mu _{3}}{\mu _{3}}\sqrt{\frac{J}{\eta _{0}}%
},  \label{solm2} \\
\zeta &=&\sqrt{\frac{\mu _{3}}{\left( \mu _{3}-\mu _{2}\right) }},
\end{eqnarray}
which is demonstrated in Fig. \ref{conv-mc}. The conversion coefficient
differs from that of the undepleted coherence case $\kappa _{m}$, Eq. (\ref
{kapm}), only by a numerical factor $2\sqrt{\left( \mu _{3}-\mu _{2}\right)
/\mu _{3}}$.


As we can see, although the total conversion efficiency may approach unity,
the robustness of the maximum coherence scheme to the phase mismatch is
lost. One has to fulfill the phase matching conditions (like (\ref{cond1}))
with sufficient accuracy to get a large conversion. A considerable transfer
of energy into the generated wave can be expected when the envelope of the
pump pulse satisfies the condition:
\begin{equation*}
1-\sqrt{f_{10}\left( t-z/c\right) }\leq \sqrt{\frac{4\mu _{2}}{\left( \mu
_{3}-\mu _{2}\right) }},
\end{equation*}
and the detuning $\delta _{3}=\delta _{3}^{0}+\delta _{3}^{\prime }$ differs
from the value $\delta _{3}^{0}$, given by condition (\ref{cond1}), by not
more than
\begin{equation*}
\delta _{3}^{\prime }\leq \sqrt{\frac{\mu _{2}}{\left( \mu _{3}-\mu
_{2}\right) }}\delta _{3}^{0}.
\end{equation*}
Both of these conditions require that the values of the transition coupling
constants, $\mu _{2}$ and $\mu _{3}$, be close.

Similar to the EIT regime, the adiabatic superposition does not
coincide with the state $\left| \psi _{0}(z=0)\right\rangle $, Eq.
(\ref{psi0}), over the entire medium, but will evolve in the
course of propagation, following the change of the fields (Fig.
\ref{conv-mc}). For large density-length products $Nz$ when
$\Omega _{1}\ll \Omega _{3}^{2}/\delta _{3}$, the atomic state
tends to the ground state: $|\psi _{0}\rangle \,\rightarrow
\,|1\rangle $. Thus, ''the maximum coherence'' is not maintained
throughout the light propagation path, although it is still
sufficiently large for most part of the medium. One can see from
Fig. \ref{conv-mc} that as the coherence gets smaller, the
conversion slows down, and for very large $Nz$ (hence, for small
atomic coherence), the generated intensity only slowly approaches
its maximum.

Here again we encounter an apparent contradiction related to the limited
validity of the adiabatic approximation. The atoms do not return to the
initial state despite the overall energy conservation. Now the preparation
energy for the atomic coherence is extracted as well. Under the conditions
discussed here, i.e. large photon number as compared to number of atoms,
this is however a small effect. A resolution of this problem should be found
when non-adiabatic corrections are taken into account, which will be the
subject of further investigations.


\subsection{Conventional nonlinear optics (weak excitation)}


Finally, we consider the regime corresponding to conventional nonlinear
optics. This regime takes place at weak excitation when both detunings $%
\delta _{2}$ and $\delta _{3}$ are much larger than all Rabi frequencies.
Then we have $\lambda \approx \Omega _{10}^{2}/\delta _{2}$ and $\left| \psi
_{0}\right\rangle \approx \left| 1\right\rangle $ for all atoms in the
medium.

The coefficients $A_{1}$ and $a_{m}$ do not depend on laser intensities in
this case:
\begin{eqnarray}
A_{1} &\simeq &q\delta _{2}\delta _{3}+\mu _{1}\delta _{3}-\mu _{3}\delta
_{2},  \label{A1n} \\
a_{0} &\simeq &\delta _{2}\delta _{3},  \label{a0n} \\
a_{1} &=&2q\left( \delta _{2}+\delta _{3}\right) +\left( \mu _{1}+\mu
_{2}-\mu _{3}\right) .  \label{a1n}
\end{eqnarray}

Since $a_{1}\min \left( \eta _{10},\eta _{20}\right) \ll a_{0}$, the
solution Eq. (\ref{sol00}) can be well approximated by
\begin{eqnarray}
J &=&J_{1}\text{sn}^{2}\left[ \kappa _{n}z;\sqrt{J_{1}/J_{2}}\right] ,
\label{soln1} \\
\kappa _{n} &=&\frac{N}{2}\frac{\sqrt{\mu _{1}\mu _{2}\mu _{3}J_{2}}}{\delta
_{2}\delta _{3}},  \label{kapn1}
\end{eqnarray}
with $J_{2,1}$ determined from Eqs. (\ref{J21}), (\ref{A1n}). This formula
is valid for any relation between $\eta _{10}$ and $\eta _{20}$.

When the phase mismatch is compensated: $A_{1}\approx 0$, which is made by
proper tuning:
\begin{equation}
q=\frac{\mu _{3}}{\delta _{3}}-\frac{\mu _{1}}{\delta _{2}},  \label{qn}
\end{equation}
the roots $J_{2,1}$ are: $J_{1}=\min \left( \eta _{10},\eta _{20}\right) $, $%
J_{2}=\max \left( \eta _{10},\eta _{20}\right) $. Therefore, the optimum
conversion can be realized also in the case of the off-resonant nonlinear
optics. In particular, for $\eta _{10}\approx \eta _{20}$ the complete
transfer of energy to the generated wave occurs:
\begin{equation*}
J=\eta _{0}\text{sn}^{2}\left[ \kappa _{n}z;p\rightarrow 1\right] \approx
\eta _{0}\tanh ^{2}\left( \kappa _{n}z\right) ,
\end{equation*}
leading to perfect total conversion efficiency, $W\approx 1$. However, the
nonlinear conversion coefficient $\kappa _{n}$ is much smaller in this case
than in the EIT and maximum coherence regimes ($\kappa _{e}$ and $\kappa
_{m} $, Eqs. (\ref{kape}) and (\ref{kapm}), respectively). Therefore, the
energy transfer is accomplished for much larger density-length products $Nz$.

At last, we note that if the phase mismatch is not compensated, we have $%
B_{1}\gg \left( \eta _{10}+\eta _{20}\right) $ and the solution is reduced
to traditional formula of nonlinear optics \cite{boyd}:
\begin{equation}
J=N^{2}\frac{\mu _{1}\mu _{2}\mu _{3}\eta _{10}\eta _{20}}{4\delta
_{2}^{2}\delta _{3}^{2}}\frac{\sin ^{2}\left( \Delta k^{\prime }z/2\right) }{%
\left( \Delta k^{\prime }/2\right) ^{2}},
\end{equation}
with the total phase mismatch $\Delta k^{\prime }=\Delta k-N\mu _{1}/2\delta
_{2}+N\mu _{3}/2\delta _{3}$.


\section{Summary}


Resonant optical processes based on atomic coherence effects such as EIT or
maximum coherence allow for a maximum nonlinear conversion of photons within
a much smaller density-length product than possible in schemes of
conventional off-resonant nonlinear optics. Since degrading mechanisms such
as phase-mismatch and absorption scale with the same density length product,
EIT or maximum coherence can thus substantially reduce the requirements for
efficient nonlinear optics. As a measure for the efficiency of using atoms
for nonlinear conversion processes, one can introduce the ratio of the
number of converted photons $n\sim \mathcal{A}\tau _{j}\bar{\eta}_{j0}$, to
the number of atoms $N_{at}=N\mathcal{A}l\sim N\mathcal{A}/\kappa $ needed ($%
\mathcal{A}$ is a cross-sectional area and $\tau _{j}$ is the duration of
the pulse). This figure of merit $n/N_{at}$ is largest for the EIT scheme: $%
n/N_{at}=\tau _{1}\Omega _{10}$, considerably smaller for the maximum
coherence method: $n/N_{at}=\tau _{2}\Omega _{20}\left( \Omega _{20}/\delta
_{3}\right) $ (with $\Omega _{20}/\delta _{3}\ll 1$), and is very small for
conventional nonlinear optics: $n/N_{at}=\tau _{2}\Omega _{20}\left( \Omega
_{10}\Omega _{20}/\delta _{3}\delta _{2}\right) $ (with $\Omega _{20}/\delta
_{3}\ll 1$ and $\Omega _{10}/\delta _{2}\ll 1$).

In previous theoretical studies of EIT- and maximum-coherence based
nonlinear optics, undepleted drive fields or a constant coherence where
assumed. If one takes into account the resources to maintain the drive field
or the constant coherence, the overall efficiency of the processes is in
this limit tiny. We have shown in the present paper that when this
restriction is lifted and coherence- or transparency maintaining fields with
comparable intensities to the pump fields are considered, it is possible to
achieve also a maximum overall conversion. To study the conditions for this,
we have derived analytical solutions for the pulse interaction in the
adiabatic limit using an approach that maps the propagation problem to that
of a nonlinear pendulum. Under certain conditions this problem could be
explicitly integrated allowing for a simple discussion of the physical
processes involved.

We have found that when the coherence- or transparency maintaining
fields are depleted, the atomic state does not remain constant but
evolves along the propagation path following the change of the
fields. Therefore, the conversion process cannot be associated
with maximum coherence or EIT in traditional sense alone.
Nevertheless, the main features of EIT, - reduced linear
absorption and refraction, and enhanced nonlinearity are still
present in the latter mechanism. Although in the ''maximum
coherence'' mechanism the robustness to the phase mismatch is
lost, the nonlinear conversion coefficient remains much larger
than in conventional nonlinear optics. Thus, for both mechanisms,
a complete conversion can be achieved within a small
density-length product.

We have also encountered several limitations of the strong adiabatic
assumption used in the Hamiltonian approach. An extension of the approach to
take into account non-adiabatic corrections and thus such important effects
as group delay is currently under investigation and corresponding results
will be presented elsewhere.

The work of E.A. K. was supported by Alexander von Humboldt Foundation. We
would like to thank R. Unanyan for bringing to our attention the Hamiltonian
formalism in nonlinear optics and the relevant publications \cite{mel79,kryz}%
, and him and K. Bergmann for many useful discussions. E.A. K. would also
like to thank K. Bergmann for the hospitality and his continuous support.


\end{document}